# Comment: Monitoring Networked Applications With Incremental Quantile Estimation

**Bin Yu**

First of all, we would like to thank the authors for their timely paper focusing on the important area of streaming data. Recently, much attention has been paid by the statistics community to the high dimensionality or massiveness of data in the information technology age. However, streaming data represent the other important feature of the IT age, the high rate of data. Both high dimensionality and high rate require fast computation, but the real-time constraint on streaming data forces its computation to be a magnitude faster than that of the off-line or batch mode of massive data. As a result, in the absence of supercomputers, the algorithms for streaming data have to be very simple to be effective.

Chambers et al. deal with streaming data for computer system monitoring. Streaming data arise also in many other fields of science and engineering, such as astronomy, geoscience and sensor networks. Chambers et al. devise a simple and practical algorithm for updating quantiles to be used to monitor the reliability of a large system based on streamed data. Stationarity is implicitly assumed since one could argue that a good computer system should be more or less stable over time until the system is updated.

A desirable add-on to the estimated quantile of Chambers et al. is a measure of uncertainty which in the i.i.d. case is trivial because of the relationship between the variance and mean of a binomial random variable. However, it is hard to imagine that a computer system follows an i.i.d. process. The real-time constraint could make the pursuit of an uncertainty measure harder than the quantile estimation itself.

For a natural environment to be monitored by a sensor network, the variable of interest (say, temperature) is most likely to be changing over time and hence nonstationary. Fortunately, there is an easy extension of the Chambers et al. algorithm to the nonstationary case. Because we can build the CDF and therefore the quantiles based on a moving window of data, it is applicable to nonstationary data streams. However, in this case, the data have to be kept over a duration of the size of the moving window $W$, in addition to the current estimate of the CDF.

Formally, let $W$ denote the size of the moving time window which is application-specific to guarantee some stationarity of the variable within the window. Let $O$ denote the initial block of (old) data to be removed when new data come in, $K$ the data block kept and $N$ the new block to be taken into account: $|W| = |O| + |K|$ and $|O| = |N|$.

Since the current empirical count of observations less than any $x$ is a summation of the indicator function of the interval $(-\infty, x]$ over the current block of data (over $K$ and $N$), it can be obtained by using the last empirical count and the summation over the old block:

$$\sum_{t \in \text{current block}} I_{\{X_t \leq x\}}$$
$$= \sum_{t \in K} I_{\{X_t \leq x\}} + \sum_{t \in N} I_{\{X_t \leq x\}}$$
$$= \sum_{t \in K} I_{\{X_t \leq x\}} + \sum_{t \in N} I_{\{X_t \leq x\}}$$
$$\quad + \sum_{t \in O} I_{\{X_t \leq x\}} - \sum_{t \in O} I_{\{X_t \leq x\}}$$
$$= \sum_{t \in O} I_{\{X_t \leq x\}} + \sum_{t \in K} I_{\{X_t \leq x\}}$$
$$\quad + \sum_{t \in N} I_{\{X_t \leq x\}} - \sum_{t \in O} I_{\{X_t \leq x\}}$$

*Bin Yu is Professor, Department of Statistics, University of California, Berkeley, Berkeley, California 94720, USA e-mail: binyu@stat.berkeley.edu*

$$= \sum_{t \in \text{previous block}} I_{\{X_t \leq x\}}$$
$$+ \sum_{t \in N} I_{\{X_t \leq x\}} - \sum_{t \in O} I_{\{X_t \leq x\}}.$$

With proper scaling and weighting, the empirical CDF for the current block can be easily updated based on the empirical CDF of the previous block, provided that the $O$-block data are kept and made available to the updating algorithm. Thus this obvious modification makes Chambers et al.'s algorithm applicable to the nonstationary case.

When the data stream is stationary, the proposed method keeps only a CDF; hence it is a form of compression. The updating data block in both the stationary and nonstationary cases might still be too expensive to communicate in situations such as sensor networks where communication is more costly than computation in terms of battery consumption. For nonstationary data, data are kept in a moving window in addition to the current CDF. If the data rate and volume are large, even a moving window of data might be too much. Hence one should compress them. It would be best if the updating could be done directly on compressed data without decompressing. This calls for an interaction of statistical analysis with data compression algorithms. Moreover, if lossy compression has to be carried out, one should allocate more bits to the tails of the distribution because the extreme quantiles are monitored for potential system anomalies. It would be interesting further research to design a bit allocation algorithm and a compression scheme to go with the quantile updating method in Chambers et al. Natural questions are: What is the objective function for bit allocation? How should it be combined with the goal of statistical estimation? What forms of codes should be designed for data compression so that they can easily interact with the CDF or quantile updating algorithm?

We would like to finish with the message that the interaction of statistical analysis with data compression algorithms is indispensable for successful and timely information extraction from high-dimensional and high-rate IT data. Although there are works to address this issue (e.g., Braverman et al., 2003, and Jörnsten et al., 2003), much more needs to be done and especially so in the streaming data context.